\documentclass[prd,preprintnumbers,floatfix,aps,notitlepage,nofootinbib,twocolumn,showpacs,amssymb]{revtex4-2}
\usepackage{amsmath,amsfonts,bm}
\usepackage{graphicx}
\usepackage{cancel}
\usepackage[colorlinks, linkcolor={red},citecolor={blue}]{hyperref}
\usepackage{xcolor}
\usepackage{color}

\newcommand{\rs}{r_{\rm s}}

\newcommand{\red}[1]{{\color{red}#1}}
\newcommand{\SA}[1]{\red{#1}}
\begin{document}
\preprint{YITP-24-122
}
\title{Analytic models for gravitational collapse}
\author{Sinya Aoki}
\email[]{saoki@yukawa.kyoto-u.ac.jp}
\affiliation{Center for Gravitational Physics and Quantum Information,
	Yukawa Institute for Theoretical Physics, Kyoto University,
	Kitashirakawa Oiwakecho, Sakyo-Ku, Kyoto 606-8502, Japan}
	
\author{Jorge Ovalle}
\email[]{corresponding author: jorge.ovalle@physics.slu.cz}
\affiliation{Research Centre for Theoretical Physics and Astrophysics,
	Institute of Physics, Silesian University in Opava, CZ-746 01 Opava, Czech Republic.}

\begin{abstract}
	We present two analytical models of gravitational collapse toward the Schwarzschild black hole, starting from the interior of the revisited Schwarzschild solution recently reported in [Phys. Rev. D 109, 104032 (2024)]. Both models satisfy some energy conditions at all times as long as the collapse is slower than some limit.     
	While a singularity of the Schwarzschild black hole at the origin ($R_{\mu\nu\alpha\beta}R^{\mu\nu\alpha\beta}\sim r^{-6}$) forms immediately after the start of the collapse in one model,
	such a singularity never appear at finite time during the collapse (except $t\to\infty$) in the other model. The scheme used shows great potential for studying in detail the appearance of singularities in general relativity. 
\end{abstract} 
\maketitle
%
%
%
\section{Introduction}
The Schwarzschild black hole (BH) solution is the simplest and most widely used BH solution in general relativity (GR). It is sourced by the simplest possible gravitational system: a point-like mass ${\cal M}$, giving rise to a physical singularity, in accordance with Penrose's singularity theorem~\cite{Penrose:1964wq}. Regarding this singularity, in a recent paper~\cite{Ovalle:2024wtv} it was address the question of whether there is alternative in GR beyond the point-like mass ${\cal M}$ as a source for the exterior region of the Schwarzschild BH, under the following conditions:
\begin{itemize}
	\item without introducing any new parameter beyond ${\cal M}$, namely, no primary hairs.
	
	\item without allowing any form of exotic matter.
\end{itemize}
As a consequence, it was found a large set of new solutions that describe the inner BH region before all form of energy encoded in the total mass ${\cal M}$ has collapsed into the singularity. Among the most attractive characteristics of these solutions, we highlight: (i) the continuity of space-time on the horizon, without the need to impose any geometric structure (thin shell) on it, (ii) tidal forces are finite everywhere for (integrable) singular solutions, and (iii) there are simple regular solutions which might be an alternative to the Schwarzschild BH as the final stage of gravitational collapse.

The importance of these solutions, apart from being new and simple, lies in the fact that they are potentially useful for studying gravitational collapse in detail: If we inforce the weak cosmic censorship conjecture~\cite{Penrose:1969pc} to avoid naked singularities, the event horizon must form before the central singularity appears. This means that part of the total mass ${\cal M}$, enveloped by the event horizon, is still on the way to the singularity. This is precisely the scenario described by the (integrable) singular solutions reported in Ref.~\cite{Ovalle:2024wtv}. This (incomplete) set of solutions shows, in a quite explicit way, the immense diversity that is possible in the interior region of the Schwarzschild BH. This is tremendously attractive if we want to study the behavior of gravity in environments of extreme curvature, such as the formation of BHs and the very early universe~\cite{Casadio:2024fol}, but especially the formation of singularities. All the above  naturally leads to wonder about details on gravitational collapse of the aforementioned solutions. This will be precisely the main goal of this work.

The paper is organized as follows: in Sec.~\ref{sec2}, we briefly review how to explore the interior of a spherically symmetric static BH by using integrable singularities, then in Sec.~\ref{sec3} we present the analytic model for the gravitational collapse, showing in detail two specific models; finally, we summarize our conclusions in Sec.~\ref{con}.


\section{Inside the black hole}
\label{sec2}
\noindent In order to be as self-contained as possible, we will now briefly outline how to explore the interior of a spherically symmetric static BH (for all details see Ref.~\cite{Ovalle:2024wtv}). Since we are interested in exploring the Schwarzschild interior, we will begin by recalling the original metric~\cite{Schwarzschild:1916uq}, given by
\begin{equation}
	\label{Schwarzschild}
	ds^{2}
	=
	-e^{\Phi(r)}\left[1-\frac{2m(r)}{r}\right]\,dt^{2}+\frac{dr^2}{1-\frac{2m(r)}{r}}+r^2d\Omega^2
	\ ,
\end{equation}
where
\begin{eqnarray}
	\label{KScond}
	&&\Phi(r)=0\ ;\\
	&&m(r)={\cal M}\ ;\,\,\,\,\,\,0<\,r\leq\infty\ .
	\label{mcond}
\end{eqnarray}
The line-element~\eqref{Schwarzschild}, under the restrictions~\eqref{KScond} and~\eqref{mcond}, has no other free parameter beyond its total mass ${\cal M}$, a point-like mass at the center $r=0$, producing a physical singularity. There is also a coordinate singularity at $r=2\,{\cal M}\equiv\,h$, which indicates the event horizon~\cite{Eddington:1924pmh,Lemaitre:1933gd,Finkelstein:1958zz,Kruskal:1959vx,Szekeres:1960gm}. On the other hand, since we want to explore a possible Schwarzschild interior beyond the singularity at $r=0$, we will change the condition~\eqref{mcond} by
\begin{equation}
	m(r)={\cal M}\ ,\,\,\,\,\,r\geq\,h\ ,
\end{equation}
where 
\begin{equation}
	\label{cond1}
	{\cal M}\equiv\,m(r)\rvert_{r=h}=\frac{h}{2}
\end{equation} 
stands for the total mass of the BH and $h$ its event horizon. Therefore, instead of~\eqref{Schwarzschild}, we have the line-element
\begin{equation}
	\label{mtransform}
	ds^2=\left\{
	\begin{array}{l}
		-\left[1-\frac{2m}{r}\right]dt^{2}+\frac{dr^2}{1-\frac{2m}{r}}+r^2d\Omega^2\ ;\,0<\,r\leq\,h\ ,
		\\
		\\
		-\left[1-\frac{2{\cal M}}{r}\right]dt^{2}+\frac{dr^2}{1-\frac{2{\cal M}}{r}}+r^2d\Omega^2 \ ;\,\,\,\,r\geq\,h\ ,
	\end{array}
	\right.
\end{equation}
where $m(r)$ is an unknown mass function associated with a Lagrangian ${\cal L}_{\rm M}\neq\,0$ which represents ordinary matter only, i.e., no electric charge, no scalar field or any other interaction\footnote{{Following Ref.~\cite{Maeda:2022lsm} for $n=4$, we can construct a BH with Schwarzschild exterior. However, we will need an additional parameter besides ${\cal M}$ for the interior $r\leq\,h$.}}. Therefore, our theory will be pure general relativity. Hence,
\begin{equation}
	\label{action}
	S=\int\left[\frac{R}{2\,\kappa}+{\cal L}_{\rm M}\right]\sqrt{-g}\,d^4x
	\ ,
\end{equation}
with $\kappa=8\,\pi\,G_{\rm N}$, $c=1$ and $R$ the scalar curvature. Notice that Eqs.~\eqref{mtransform} and~\eqref{action} means that ${\cal L}_{\rm M}=0$ for $r>h$ only, and therefore for the interior $r\leq\,h$ Einstein field equations become  
\begin{eqnarray} 
	\label{sources}
	\rho=\frac{2{m}'}{\kappa\,r^2}\ ,\,\,\,\,\,\,p_r=-\frac{2{m}'}{\kappa\,r^2}\ ,\,\,\,\,\,\,p_\theta=-
	\frac{{m}''}{\kappa\,r}\ .
\end{eqnarray}
where the energy-momentum tensor 
\begin{eqnarray}
	\label{emt}
	T^{\mu}_{\,\,\,\nu}={\rm diag}[\,p_r,-\rho,\,p_\theta,\,p_\theta]\ ,
\end{eqnarray}
contains an energy density $\rho$, radial pressure $p_r$ and transverse pressure $p_\theta$. We see that the particular form of {$T^\mu_{\,\,\nu}$} in~\eqref{emt} means the radial and temporal coordinates exchange roles for $r\leq\,h$. Finally, notice that Einstein equations in~\eqref{sources} are linear in the mass function ${m}(r)$. Hence, any solution $m(r)$ of~\eqref{sources} can be coupled with a second one $\hat{m}(r)$ to produce a new solution $\bar{m}(r)$ as 
\begin{eqnarray}
	\label{gdks}
	m(r)
	&\rightarrow & \bar{m}(r) = m(r)+\hat{m}(r)
	\ ,
\end{eqnarray}
which represents a trivial case of the so-called gravitational decoupling~\cite{Ovalle:2017fgl,Ovalle:2019qyi}. 
\par
Regarding the Bianchi identity for the region inside the BH, it leads to $\nabla_\mu\,{T}^{\mu\nu}=0$, which can be expressed as
\begin{eqnarray}
	\label{con111}
	\rho'=
	-\frac{2}{r}\left(p_\theta-p_r\right)
	\ ,
\end{eqnarray}
namely, the Tolman–Oppenheimer–Volkoff (TOV) equation for hydrostatic equilibrium.
Since the energy density $\rho$ is expected to decrease monotonically from the origin, i.e., $\rho'<0$, the expression~\eqref{con111} requires
\begin{equation}
	\label{anis}
	p_\theta>p_r
	\ .
\end{equation}
Therefore, we conclude that the anisotropic fluid experiences a pull towards the center as a consequence of negative energy
gradients $\rho'<0$ that is canceled by a gravitational repulsion caused precisely by the anisotropic
pressure. We would like to conclude by highlighting that the equilibrium (stable or not) displayed in Eq.~\eqref{con111}
does not mean the fluid element will not face the singularity, which is the endpoint of geodesics inside singular BHs.

In order to investigate the generic matter $m(r)$ inside the BH as a source for the exterior Schwarzschild solution, we need to examine the continuity of the metric~\eqref{mtransform} at the horizon $r=h$. In order to joint both regions in Eq.~\eqref{mtransform}, the mass function $m(r)$ must satisfy
\begin{equation}
	\label{cond2}
	m(h)={\cal M}\ ;\,\,\,\,\,\,\,m'(h)=0\ ,
\end{equation}
where $F(h)\equiv\,F(r)\big\rvert_{r=\rs}$ for any $F(r)$. Expressions in Eq.~\eqref{cond2} are the necessary and sufficient conditions for match the interior with the Schwarzschild exterior at $r=h$. Finally, we see from Eqs.~\eqref{sources} and~\eqref{cond2} that the continuity of the mass function $m(r)$ leads to the continuity of both density and radial pressure. Hence, 
\begin{equation}
	\label{c2a}
	\rho(h)=0\ ;\,\,\,\,\,\,\,\,p_r(h)=0\ .
\end{equation}
However, the pressure $p_\theta\sim\,m''$, and therefore the energy-momentum tensor in Eq.~\eqref{emt}, are in general discontinuous, namely, $p_\theta(h)\neq\,0$ and $T^\mu_{\,\,\nu}(h)\neq\,0$. 
\par
At this stage we might wonder about the compatibility of some known BH solutions, as Hayward~\cite{Hayward:2005gi}, Bardeen~\cite{Bardeen:1968qtr} and Simpson \& Visser~\cite{Simpson:2021dyo}, with the line element~\eqref{mtransform}. In the case of these three popular solutions, all of them have two free parameters, i.e. $\{M,\,\ell\}$, which after imposing the condition~\eqref{cond2}, lead us to trivial solutions (Minkowski spacetime or Schwarzschild solution). The same occurs for a large number of known solutions, which indicates that the problem is far from trivial.

\subsection*{Integrable singularities}
\label{sec3} 
\noindent BH solutions, in general, can be grouped into two types: (i) singular, those that have a physical singularity, and (ii) regular, those that lack such singularities. Regarding regular BHs, it is well known that this type of solutions, in addition to an event horizon, has an inner (Cauchy) horizon, which turns out to be particularly problematic. This implies (related) problems such as mass inflation, instability, and eventual loss of causality~\cite{Poisson:1989zz,Poisson:1990eh} (see also Ref.~\cite{Ori:1991zz} and Refs.~\cite{Bonanno:2020fgp,Carballo-Rubio:2021bpr,Carballo-Rubio:2022kad,Franzin:2022wai,Casadio:2022ndh,Bonanno:2022jjp,Casadio:2023iqt,McMaken:2023uue} for a recent study). Fortunately, there is a third type of solutions which we can place between the two aforementioned families: integrable BHs. These are still singulars, but unlike the previous ones, they are characterized by a singularity in the curvature $R$ that occurs at most as
\begin{equation}
	\label{inte-sing}
	R\sim\,r^{-2}\ .
\end{equation}
The main feature of BHs with this kind of singularities is that tidal forces remains finite everywhere~\cite{Lukash:2013ts}, they also 
have a well-defined finite mass, and (in general) lack Cauchy horizons. A large family of BH solutions with integrable singularities can be found~\cite{Ovalle:2024wtv} by the mass function 
\begin{equation}
	\label{M}
	m(r)=M-\frac{Q^2}{2\,r}+\frac{1}{2}\sum_{n=0}^{{\infty}}\,\frac{C_n\,r^{n+1}}{(n+1)(n+2)}\ ;\,\,\,\,\,r\leq\,h\ ,
\end{equation}
where $\{M,\,Q\}$ are two integration constants that are naturally identified with the mass of the Schwarzschild solution and a charge for the Reissner-Nordstr\"{o}m (RN) geometry, respectively. However, since we are interested in pure general relativity, as we show by the action~\eqref{action}, we discard $Q$ as an electric charge.  In any case, since it is known that the RN geometry contains a Cauchy horizon, we impose $Q=0$. Regarding the generic integrable solution~\eqref{M}, first of all notice that the infinite series in~\eqref{M} converges as soon as we impose the condition~\eqref{cond1}. However, whether this infinite series represents an analytic function is something we do not know yet. Secondly, we can see that the Schwarzschild BH solution is given by 
\begin{eqnarray}
	\label{Schw-limit}
	&&M={\cal M}\neq\,0\ ,\nonumber
	\\ 
	\\
	&&Q=C_n=0\ ,\nonumber
\end{eqnarray}
for all $n$ in Eq.~\eqref{M}. Finally, following the Ref.~\cite{Ovalle:2024wtv}, solutions determined only by the total mass ${\cal M}$ of the configuration imply no primary hairs, namely,
\begin{eqnarray}
	\label{Schw-limit2}
	M=Q=0\ .
\end{eqnarray}
Notice that the condition~\eqref{Schw-limit2} is not contained in the Schwarzschild limit~\eqref{Schw-limit}, and therefore it seems to be in conflict with the Schwarzschild solution~\eqref{Schwarzschild}. However, as it was proved in Ref.~\cite{Ovalle:2024wtv}, the condition~\eqref{Schw-limit2} is perfectly compatible with the Schwarzschild solution, and indeed it is the key for finding the set of solutions corresponding to the revisited Schwarzschild interior. 
\subsection*{The simplest solution with {$T^\mu_{\,\,\nu}\rvert_{r=h}=0$}} 
\noindent Following the procedure previously described, we can construct a large number of interior solutions, perfectly compatible with the exterior of the Schwarzschild BH. In particular, we can construct a complete family of BH solutions that satisfy the continuity of {$T^\mu_{\,\,\nu}$} on the horizon. This can be accomplished by demanding, in addition to conditions~\eqref{cond2}, continuity of the second derivative of the mass function at the horizon, i,e.,
	\begin{equation}
		\label{cond3}
		m''(r)\rvert_{r=h}=0\ .
	\end{equation}
	Among all these solutions, we highlight the simplest of all, given by the mass function
\begin{equation}
	\label{M2}
	m(r)=r-\frac{r^3}{h^2}+\frac{r^4}{2\,h^3}\ ,
\end{equation}
which used in the metric~\eqref{mtransform} yields
\begin{eqnarray}
	\label{sol2}
	ds^{2}
	= 
	&&\left(1-\frac{2\,r^2}{h^2}+\frac{r^3}{h^3}\right)\,dt^{2}-\frac{dr^2}{	\left(1-\frac{2\,r^2}{h^2}+\frac{r^3}{h^3}\right)}\nonumber\\
	&&+\,r^2\,d\Omega^2\ ;\,\,\,\,\,\,\,\,\,\,\,\,\,\,\,\,\,\,\,\,\,\,\,\,\,\,\,\,\,\,\,\,\,\,\,\,\,\,\,\,\,\,\,\,\,\,\,\,\,\,\,\,\,\,\,\,\,\,\,\, r\leq\,h	\ .
\end{eqnarray}
The nontrivial source for the metric~\eqref{sol2}, which also gene\-rates the outer Schwarzhchild BH, is given by
\begin{eqnarray}
	\label{sources2}
	&&	\kappa\rho=-\kappa\,p_r=\frac{2}{r^2\,h^3}\left(h-r\right)^2(h+2\,r)\ ,\nonumber\\
	&&	\kappa\,p_\theta=\frac{6}{h^3}\left(h-r\right)\ ,
\end{eqnarray} 
 \begin{figure}
	\centering
	\includegraphics[width=0.46\textwidth]{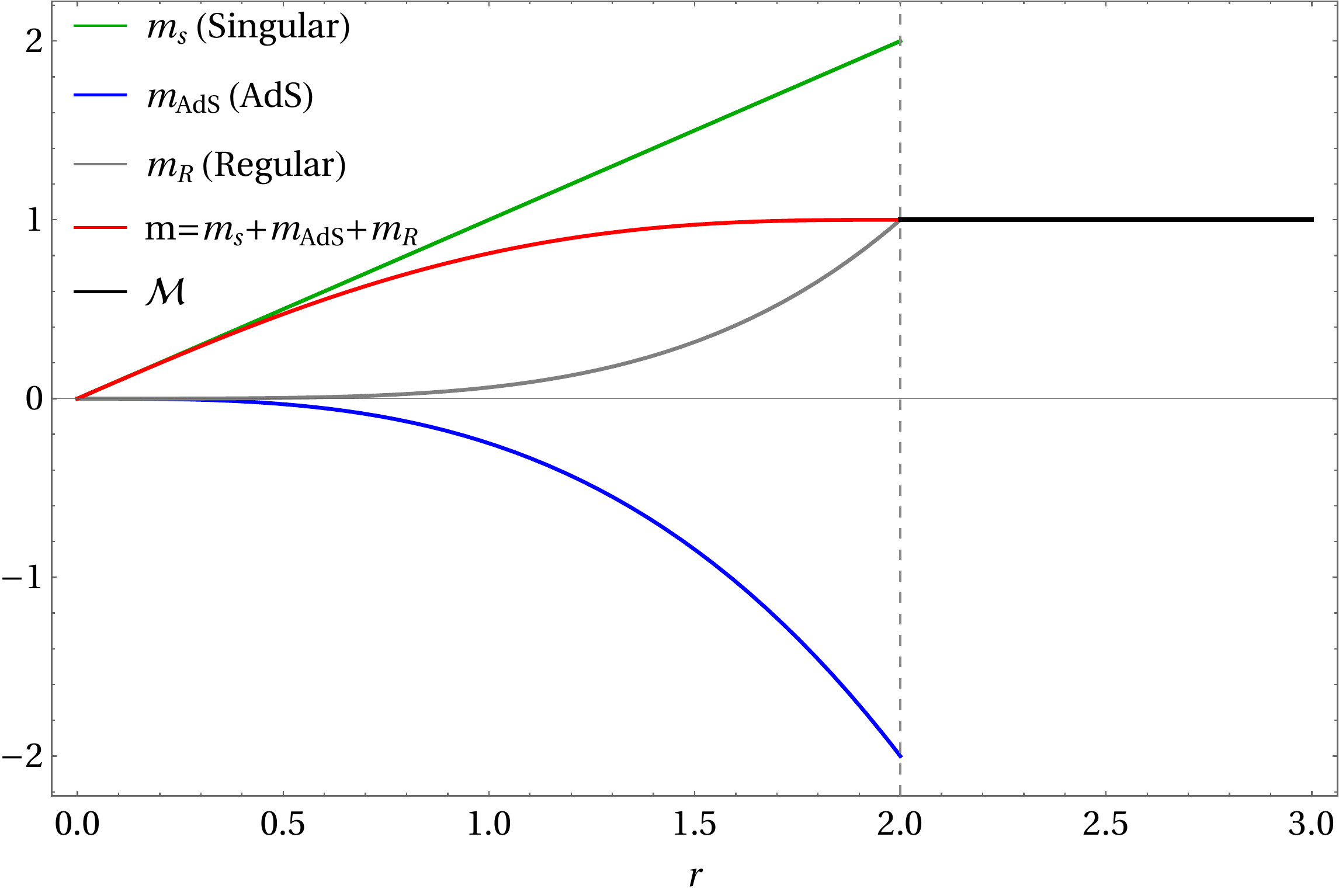}
	\caption{Mass function $m(r)$ in Eq.~\eqref{M2}. It is formed by the superposition of three different configurations. Its dynamical version is displayed in Fig.~\ref{fig2}. We take ${\cal M}=1$.}
	\label{fig3}
\end{figure}
with curvature 
\begin{equation}
	\label{Rsin2}
	R=\frac{4}{r^2}\left[1+\frac{5\,r^3}{h^3}-\frac{6\,r^2}{h^2}\right]\ ;\,\,\,\,\,\,r<h\ .
\end{equation}
As we see through the scalar~\eqref{Rsin2}, the Schwarzschild interior~\eqref{sol2} is formed by the superposition of a singular solution, an anti-de Siter, and a regular solution, as shown in Fig.~\ref{fig3}. We would like to conclude by highlighting that the inner region can extend even beyond what is described in Ref.~\cite{Ovalle:2024wtv}, thus offering many possibilities. This will be particularly important for gravitational collapse models.

\section{Gravitational collapse}
\label{sec3}
\par
\noindent In order to study the gravitational collapse associated with the inner geometry of a BH, let us perform an appropriate coordinated transformation in the line element~\eqref{Schwarzschild} [with $\Phi(r)=0$ and $m(r)\neq{\cal M}$], namely, 
\begin{equation}
	\label{EF-transf}
	t\,\rightarrow\,\bar{t}=t-\int\,\frac{u}{(1+u)}\,dr, \ u:= -{2m(r)\over r} .
\end{equation}  
which allows to find its Eddington-Finkelstein form as (dropping bars)
\begin{equation}
	\label{EF-metric}
	ds^{2}
	=
	-(1+u)\,dt^{2}-2u\,dt\,dr+\left(1-u\right)dr^2+r^2\,d\Omega^2\ .
\end{equation}
The metric~\eqref{EF-metric}, contrary to~\eqref{Schwarzschild}, has no (coordinated) singularities at $r=h$ ($u=-1$). We can now explore the Schwarzschild interior, for any of the alternatives in Ref.~\cite{Ovalle:2024wtv} (and eventual extensions), in much more convenient coordinates.
Since a constant $t$ hyper-surface is  always space-like even inside the horizon ($r\le h$) ,
we can regard $t$ as an appropriate time coordinate.

Since we are interested in building simple and exact analytical models of the collapse, we will begin by promoting the mass function to a completely generic dynamic form, that is,
\begin{equation}
	m(r)\rightarrow\,m(r,t), \  u(r)\to u(r,t) = -{2m(r,t)\over r}.
\end{equation}
In this case the Einstein tensor reads~\cite{Aoki:2020nzm,Aoki:2022gez}
\begin{eqnarray}
&&G^0_{\,\,0}=-\frac{2m'}{r^2}:=-\kappa A\ ,\label{G00}\\
&&G^1_{\,\,1}=-\frac{2}{r^2}\left(m'-2\dot{m}\right):= \kappa (-A+2B)\ ,\\
&&G^2_{\,\,2}=G^3_{\,\,3}=-\frac{1}{r}\left(m''-2\dot{m'}+\ddot{m}\right):=-\kappa C\ ,\\
&&G^0_{\,\,1}=-G^1_{\,\,0}=-\frac{2\dot{m}}{r^2}:= -\kappa B\label{G01}\ ,
\end{eqnarray}
where $\dot{m}\equiv\frac{\partial\,m}{\partial\,t}$, and we define $A,B,C$ for latter uses.

The corresponding energy momentum tensor, which is the type II of the Hawking Ellis classification~\cite{Hawking:1973uf,Martin-Moruno:2018eil,Maeda:2022vld}, 
is written as
\begin{eqnarray}
T^{\mu\nu} &=&\left(A-B\right) (\ell^\mu s ^\nu+s^\mu \ell^\nu)+B\ell^\mu \ell^\nu -C\sum_{i=1}^2 e^\mu_i e^\nu_i,~~~~~~
\end{eqnarray}
where
\begin{eqnarray}
\ell^\mu &=& \delta^\mu_0-\delta^\mu_1, \  
s^\mu = {1-u\over 2}\delta^\mu_0+{1+u\over 2}\delta^\mu_1,\nonumber \\
e^\mu_2 &=&{1\over r}\delta^\mu_2, \ e^\mu_3={1\over r\sin\theta} \delta^\mu_3,
\end{eqnarray}
which satisfies  $\ell^\mu\ell_\mu=s^\mu s_\mu=0$ and  $\ell^\mu s_\mu =-1$.
Using  the orthonormal basis $e^\mu_i$ which satisfies $ g_{\mu\nu} e^\mu_ie^\nu_j =\eta_{ij} ={\rm daig}(1,-1,1,1)$,
we  rewrite it as 
\begin{eqnarray}
T^{\mu\nu} &=&\left(A-{B\over 2}\right) e^\mu_1e^\nu_1+\left({3B\over 2}-A\right) e^\mu_0 e^\nu_0\nonumber \\
&-&{B\over 2} (e^\mu_0e^\nu_1+e^\mu_1e^\nu_0) -C\sum_{i=1}^2 e^\mu_i e^\nu_i,
\end{eqnarray}
where
\begin{eqnarray}
e^\mu_1 &=& {\ell^\mu+s^\mu\over \sqrt{2}}= 
{3-u\over 2\sqrt{2}}\delta^\mu_0 - {1-u\over 2\sqrt{2}}\delta^\mu_1, \nonumber \\
e^\mu_0 &=&  {-\ell^\mu+s^\mu\over \sqrt{2}}=
{3+u\over 2\sqrt{2}}\delta^\mu_1 -{1+u\over 2\sqrt{2}}\delta^\mu_0.
\end{eqnarray}

The components of the energy momentum tensor in the the orthonormal basis can be found by
\begin{equation}
	T_{ij}=e^\mu_{\, i}e^\nu_{\, j}T_{\mu\nu}\ ,
\end{equation}
which explicitly reads 
\begin{eqnarray}
		\label{or01}
	&&T_{01}=T_{01}={B\over 2}=\frac{\dot{m}}{\kappa r^2}\ ,\\
	\label{or00}
	&&T_{11}=A-{B\over 2}=\frac{2m'-\dot{m}}{\kappa r^2}\ ,\\
		\label{or11}
	&&T_{00}={3B\over 2} -A=\frac{3\dot{m}-2m'}{\kappa r^2}\ ,\\
		\label{or22}
	&&T_{22}=T_{33}=-C=-{\left(m''-2\dot{m'}+\ddot{m}\right)\over \kappa r}\ .
\end{eqnarray}
From expressions~\eqref{or01}-\eqref{or22} we find that the radiation flux $\epsilon$, energy density $\rho$, radial pressure $p_r$ and tangential pressure $p_\theta$ are given, respectively, by
\begin{eqnarray}
		\label{fluxg}
	&&\epsilon=-T^{10}=T_{01}=\frac{\dot{m}}{\kappa r^2}\ ,\\
	\label{rhog}
	&&\rho=T_{11}=-T^{1}_{\,\,1} =\frac{2m'-\dot{m}}{\kappa r^2}\ ,\\
		\label{prg}
	&&p_r=T_{00}=T^{0}_{\,\,0} =\frac{3\dot{m}-2m'}{\kappa r^2}\ ,\\
		\label{ptg}
	&&p_\theta=T^{2}_{\,\,2} =T^{3}_{\,\,3} =-{\left(m''-2\dot{m'}+\ddot{m}\right)\over \kappa r}\ .
\end{eqnarray}
Combining~\eqref{fluxg},~\eqref{rhog} and~\eqref{prg} we find the equation of state
\begin{equation}
	\label{es}
{p}_r=2\,\epsilon-\rho\ .
\end{equation}


Note that for the gravitational collapse inside a BH, the positive energy flux $+\epsilon$ is defined as that which always points towards the singularity, i.e., the future. Therefore, the energy flux $\epsilon$ passing through a given closed surface (enclosing $r=0$) will increase the mass $m$ enclosed within that surface. Hence $\dot{m}>0\rightarrow{\epsilon>0}$, in agreement with the expression~\eqref{fluxg}. Next, in order to elucidate certain details of the gravitational collapse, in particular the formation of the central singularity, we will consider some dynamical versions of the inner geometries in Ref.~\cite{Ovalle:2024wtv}.

\subsection*{Formation of the Schwarzschild vacuum: Model I}

\begin{figure}
	\centering
	\includegraphics[width=0.50\textwidth]{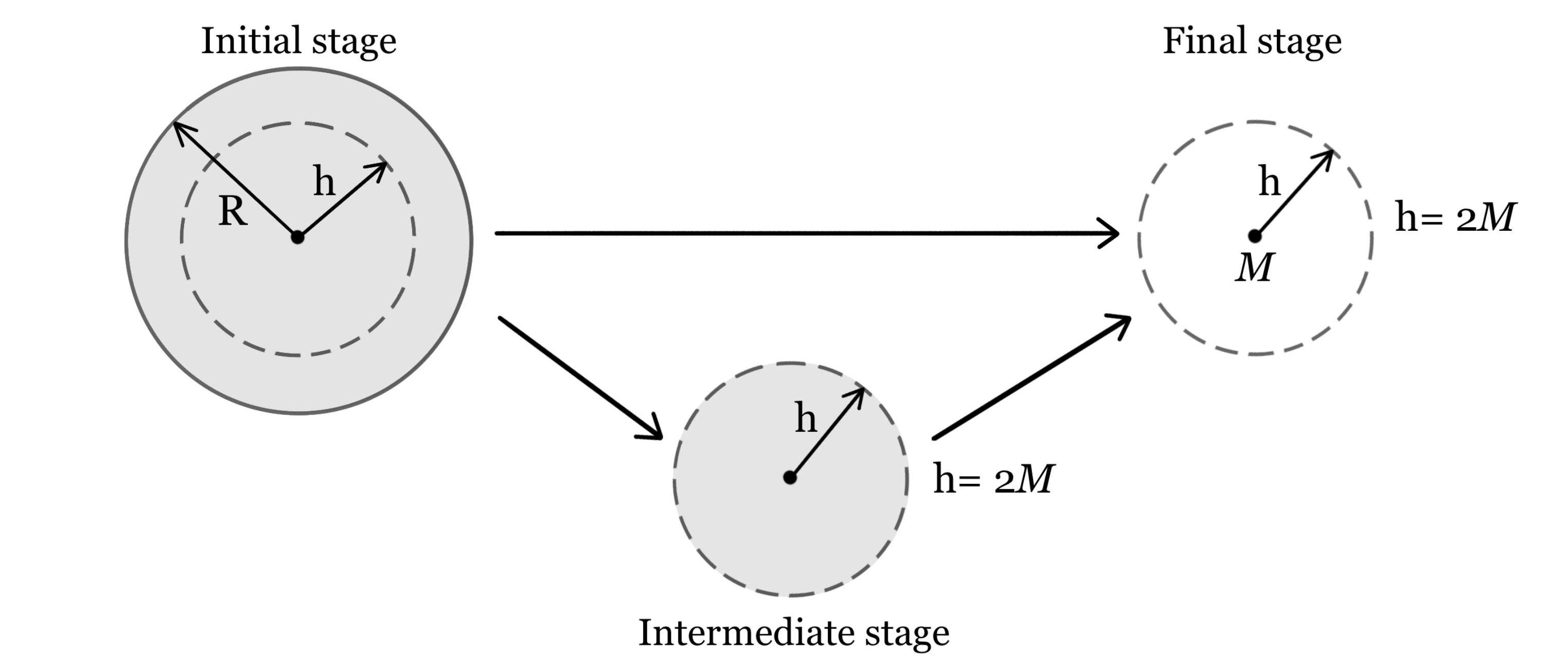}
	\caption{Gravitational collapse and formation of singularities. We know how it begins (collapsing star) an ends (singularity). The intermediate state, described by the revisited Schwarzschild BH, is exact and analytic.}
	\label{fig1}
\end{figure}

\begin{figure}
	\centering
	\includegraphics[width=0.50\textwidth]{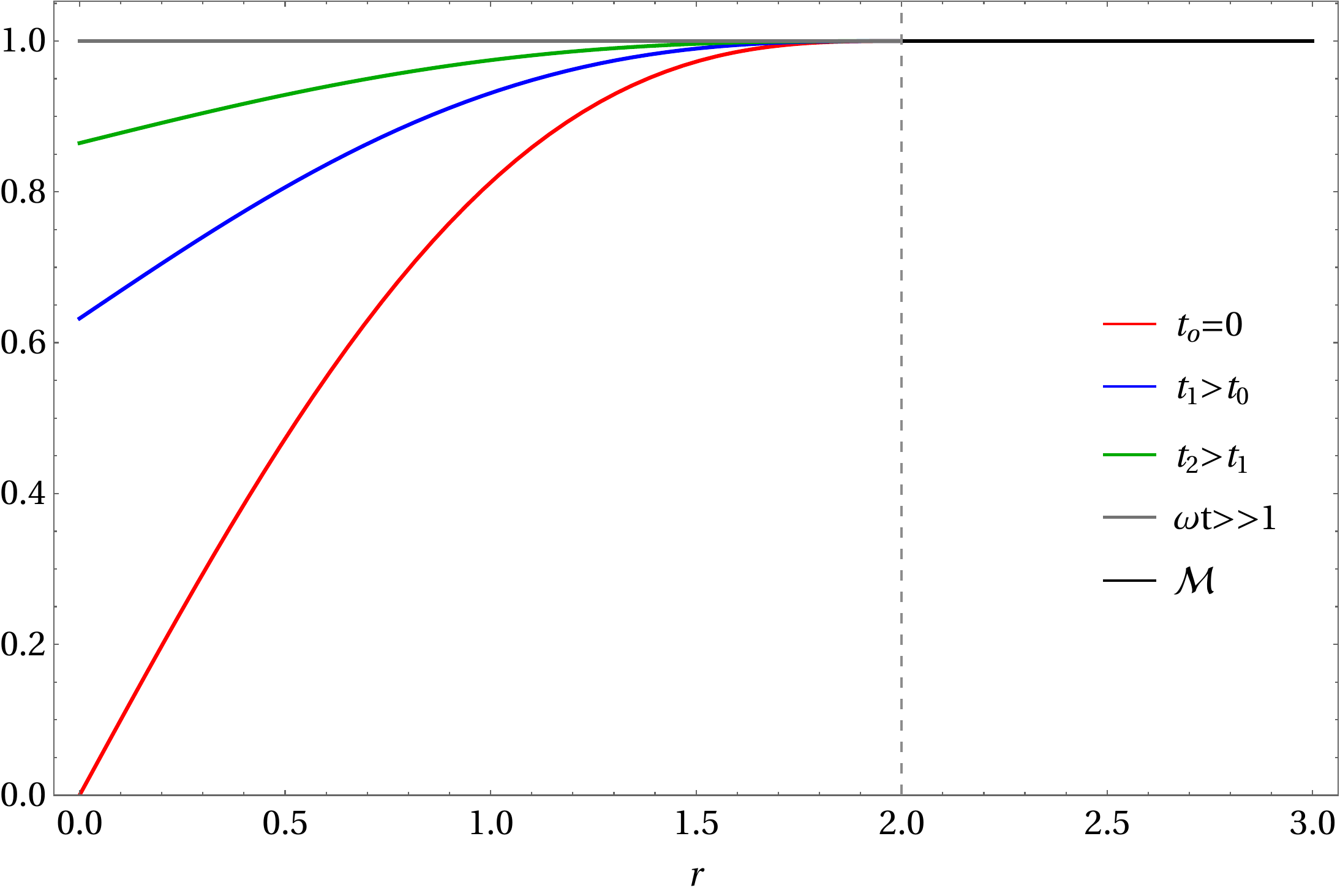}
	\caption{Mass function $m(r,t)$ in Eq.~\eqref{m(t)} for $m(r)\equiv\,m(r,0)$ in Eq.~\eqref{M2}. 
		We see the matter collapsing inside the black hole. The vacuum is formed around the Schwarzschild singularity for $\omega\,t>>1$. We take ${\cal M}=1$ and $\omega=0.1$.} 
	\label{fig2}
\end{figure}

\noindent As we pointed out in Ref.~\cite{Ovalle:2024wtv}, the Schwarzschild geometry in Eq.~\eqref{Schwarzschild} shows the final state of gravitational collapse, without giving details about this process. On the other hand, since the event horizon forms before the central singularity appears~\cite{Penrose:1969pc}, part of the total mass $\cal M$ is still on its way to the singularity. This is precisely the scenario described by the revisited Schwarzschild BH. Therefore, if we want to elucidate some aspects of the aforementioned process, the first step will be to promote these solutions to the dynamical case. To carry out this, the first step will be to specify the initial and final conditions of the collapse, namely,
\begin{equation}
	m(r,t)\vert_{t=0}=m(r)\ ;\,\,\,	m(r,t)\mid_{t\rightarrow\,\infty}={\cal M}\ .
\end{equation}
The simplest possible model satisfying the above is
\begin{equation}
	\label{m(t)}
	m(r,t)={\cal M}+\left[m(r)-{\cal M}\right]e^{-\omega\,t}\ ,
\end{equation}
where $\omega^{-1}$ is a time scale associated with the collapse, and $m(r)$ an initial configuration for collapsing matter inside BHs, as those for the revisited Schwarzschild BH. Since a large $\tau\equiv\omega^{-1}$ means a large inertia to collapse, it is reasonable to conclude that 
\begin{equation}
	\label{tau}
\tau\equiv\omega^{-1}\sim\,h\ ,
\end{equation}
that is, small for stellar BHs and very large for super massive BHs. Finally, note that 
we are building a dynamic solution whose static case has a (integrable) singularity, such as those represented in the revisited Schwarzschild BH, and schematized in Fig.~\ref{fig1}. This means that the singularity is already present and therefore our model is describing the formation of the Schwarzschild vacuum [see Eqs.~\eqref{scalars1},~\eqref{krets1}]. 
The model, although quite simple, is sufficient for our purpose: an analytic study of the last stage of gravitational collapse, that where both the horizon and the singularity have already formed, but there is still matter collapsing, as we show in Fig.~\ref{fig2}.

\par

Using the mass function \eqref{m(t)} in Eqs.~\eqref{fluxg}-\eqref{ptg}, we have
\begin{eqnarray}
		\label{fluxg2}
	&&\hspace*{-8mm}\epsilon=\omega e^{-\omega\,t}\frac{{\cal M}-m(r)}{\kappa r^2}\ \ ,\\
		\label{rhog2}
	&&\hspace*{-8mm}\rho=e^{-\omega\,t} \frac{ 2 m'(r)+\omega[m(r) -{\cal M} ]}{\kappa r^2}\ ,\\ 
		\label{ptg2}
	&&\hspace*{-8mm}p_\theta=e^{-\omega\,t}\frac{-m''(r) -2\omega m'(r) +\omega^2[ m(r)-{\cal M}]}{\kappa r}\ ,
\end{eqnarray}
where $p_r$ is given by~\eqref{es}, and $m(r)$ in Eqs.~\eqref{fluxg2}-\eqref{ptg2} represents any of the inner geometries for the revisited Schwarzschild BH. 

The next step will be to choose any of the solutions displayed in Ref.~\cite{Ovalle:2024wtv}. In the case of the simplest solution with {$T^\mu_{\,\,\nu}\rvert_{r=h}=0$}, given by Eq.~\eqref{M2}, the expressions in Eqs.~\eqref{fluxg2}-\eqref{ptg2} read
\begin{eqnarray}
	\label{fluxs1}
	&&\epsilon=\alpha \, e^{-\omega\,t}\frac{(1-x)^3(1+x)}{2\kappa r} \ , \quad \alpha:=\omega h, \ x:={r\over h} \ , \\
	\label{rhos1}
&&\rho=\frac{e^{-\omega\,t}(1-x)^2}{2\kappa r^2}\left[ 4(2x+1) -\alpha (1-x)(1+x)\right] \ , \\
		\label{pts1}
	&&p_\theta=\frac{e^{-\omega\,t} x(1-x)}{2\kappa r^2}\times  \nonumber \\
	& & \left[12 x-4\alpha(1-x)(2x+1) -\alpha^2(1-x)^2(1+x)\right] \ ,
	\end{eqnarray}
where $p_r$ is given by~\eqref{es}. As expected, expressions in Eqs.~\eqref{fluxs1}-\eqref{pts1} are reduced to those in Eq.~\eqref{sources2} for $\alpha=\omega h=0$. Also, for $t\rightarrow\,\infty$ we have the vacuum solution $\epsilon=\rho=p_r=p_t=0$ at $r\not= 0$, namely, the Schwarzschild singularity. In particular, we want to emphasize that the scalar curvature and Kretschmann scalar behave respectively as
\begin{eqnarray}
	\label{scalars1}
	&&R\sim\,\frac{2\,e^{-\omega\,t}}{r^2}\left(2-h\omega\right)+\frac{\omega\,e^{-\omega\,t}}{r}\left(8-h\omega\right)+{\cal O}\ ,\\
	\label{krets1}
	&&R_{\mu\nu\sigma\rho}R^{\mu\nu\sigma\rho}\sim\,\frac{12\,h^2}{r^6}\left(1-e^{-\omega\,t}\right)^2+{\cal O}\ .
\end{eqnarray}
that is,  
in addition to the integrable singularity ($R\sim\,r^{-2}$), the Schwarzschild singularity at $r=0$ develops as long as $t> 0$ during the collapse. 
(See Ref.~\cite{Adler:2005vn} for a similar type of gravitational collapse.)

 Regarding the fluid and its physical acceptability, 
we  have investigated the energy conditions in detail as shown in appendix~\ref{sec:EnergyCondition}, as the energy momentum tensor can not be diagonalized.
The analysis shows that the Null and Weak energy conditions (NEC and WEC) are satisfied as long as  
\begin{equation}
	\label{W-cond}
\alpha=h\omega\leq\,2\ .
\end{equation}
while the Strong energy condition (SEC) is violated except for the static case ($\alpha=0$).
The Dominant energy condition (DEC) is always violated even without time-dependence ($\alpha=0$).\footnote{A violation of the DEC has been also observed for slowly formed black holes from quantum matters\cite{Kawai:2021qdk}.}
The expression in Eq.~\eqref{W-cond} is quite significant. It establishes that for a given $h$ the collapse cannot be arbitrary. In this regard, if we introduce a collapse factor $X(t, \tau) := e^{-t/\tau}$,  we see that it must always evolve slower than its critical value, i.e.,
\begin{equation}
	\label{crit}
X(t,\tau)\leq\,X_C\equiv\,e^{-2t/h}=\,e^{-t/{\cal M}}\ .
\end{equation}
This means that super massive BHs evolve slower than stellar BHs, in agreement with the expression in Eq.~\eqref{tau}.
Finally, notice that according to~\eqref{scalars1}, only those configurations with a positive scalar curvature $R$ around $r=0$ are allowed during the collapse.

We finally  consider the energy conservation during the collapse.
As in  Ref.~\cite{Adler:2005vn} for similar models,
the total energy can be written as
\begin{eqnarray}
E =E_{\rm BH} + E_{\rm matter},\quad E_{\rm BH} = {{\cal M}\over G_N} \left(1-e^{-\omega t} \right)
\end{eqnarray}
where $G_N$ is the Newton constant. 
We evaluate the matter energy $E_{\rm matter}$ on the constant $t$ hyper-surface, which is always space-like and whose volume element is given by
$d\Sigma_\mu = -r^2dr \sin\theta d\theta d\phi \delta_\mu^t $. We then obtain 
\begin{eqnarray}
E_{\rm matter} &=&\int d\Sigma_\mu T^\mu{}_0 = -\int r^2dr \sin\theta d\theta d\phi T^0{}_0 \nonumber \\
&=& {8\pi\over \kappa}\int_0^\infty dr\, \partial_r m(r,t)={{\cal M}\over G_N}  e^{-\omega t}, 
\end{eqnarray}
so that the total energy is indeed $t$-independent as
\begin{eqnarray}
E &=&  {{\cal M}\over G_N} \left(1-e^{-\omega t} \right)+{{\cal M}\over G_N}  e^{-\omega t}= \frac{\cal M}{G_N}.
\end{eqnarray}
  
\subsection*{Formation of the Schwarzschild vacuum: Model II}
We here consider a different model of the time dependent mass function. Let us start by introducing, for any mass $m(r)$ of the revisited Schwarzschild BH, the function
\begin{equation}
	{g\left({r\over h}\right)}=\frac{m(r)}{h}\ .
\end{equation}
For the simplest case, given by Eq.~\eqref{M2}, \SA{it} reads
\begin{eqnarray}
	\label{g-fun}
	g(x) = {1 -(1-x)^3(1+x)\over 2}.
\end{eqnarray}
where $x$ is defined in Eq.~\eqref{fluxs1}. Now let us introduce the mass function
\begin{eqnarray}
m(r,t) &=& \left\{
\begin{array}{ll}
 2{\cal M} g(x(t)), & \displaystyle  x(t) \le 1,     \\
 \\
  {\cal M}, & x(t)  > 1 .     \\
\end{array}
\right.
\label{eq:m2(t)}
\end{eqnarray}
where $g$, given by~\eqref{g-fun}, has been promoted to a time dependence $g(x(t))$ by
\begin{equation}
	x\rightarrow\,x(t):=\frac{r}{h\,{e^{-\omega t}}}\ .
\end{equation}
\begin{figure}
	\centering
	\includegraphics[width=0.50\textwidth]{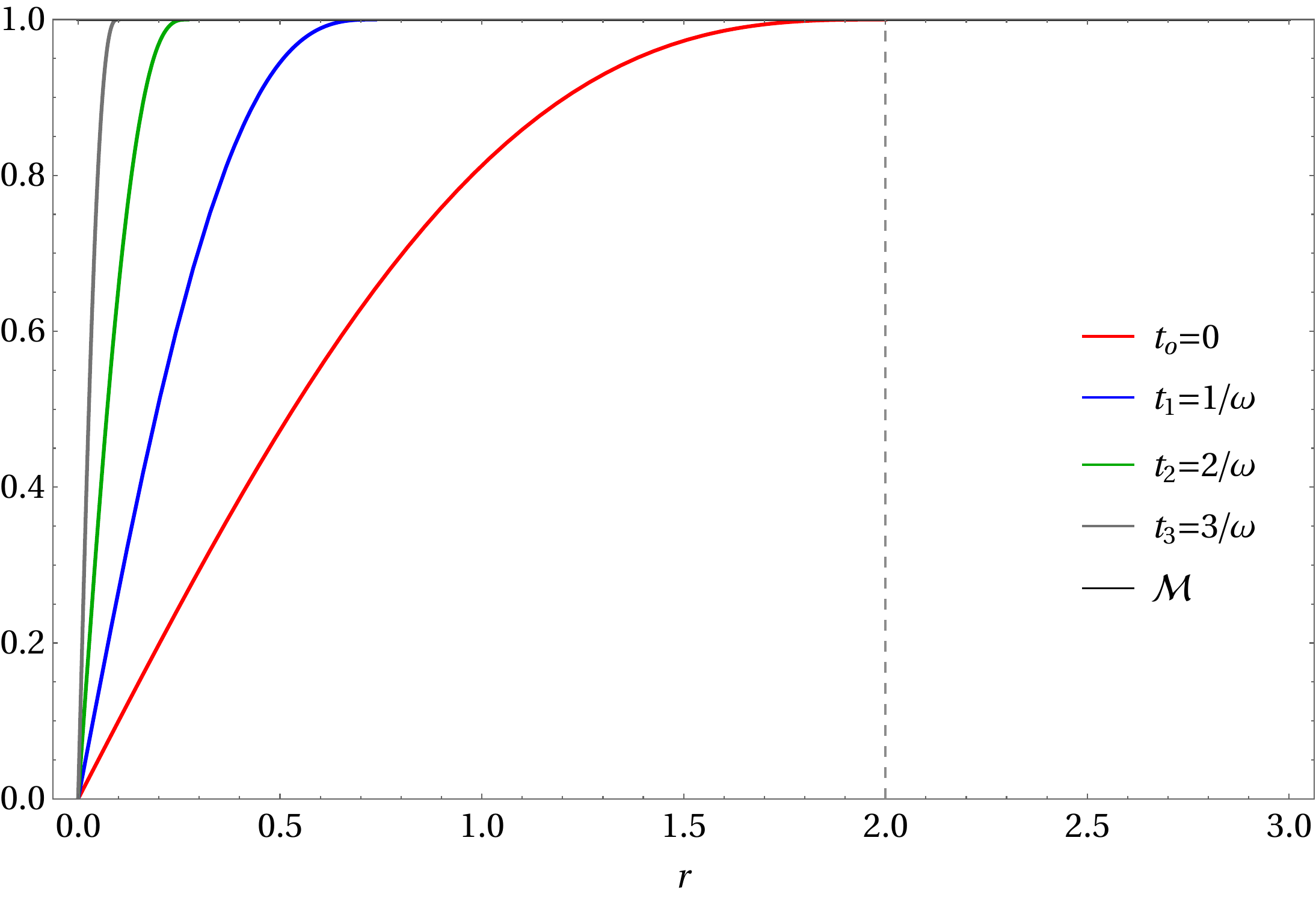}
	\caption{Mass function $m(r,t)$ in Eq.~\eqref{eq:m2(t)} for $m(r)$ in Eq.~\eqref{M2} as a function of $r$ at $t=0$ (red), $t=1/\omega$ (blue), $t=2/\omega$ (Green), and $t=3/\omega$ (gray).
	While $m(r,t)$ approaches the profile of the constant $m(r)={\cal M}$ as $t$ increases, the matter never reaches at $r=0$ for all $t$.	
	We take ${\cal M}=1$ and $\omega=0.1$.} 
	\label{fig4}
\end{figure}
We plot $m(r,t)$ as a function of $r$ for $h=2$  (${\cal M}=1$ and $\omega = 0.1$ in Fig.~\ref{fig4}.
We see that  the mass function of $m(r,t)$ approach that of the Schwarzschild BH except at $r=0$ as $t$ increases, where
$m(0,t)=0$ for all $t$. Therefore we obtain
\begin{eqnarray}
\lim_{t\to\infty} m(r,t) ={\cal M} \theta (r),
\label{eq:SBH_mass}
\end{eqnarray}
which is nothing but a mass function of the Schwarzschild BH with a regularized singularity at $r=0$ by the step function 
$\theta(r)$ satisfying $\theta(r)=1$ for $ r>0$ with $\theta(0)=0$~\cite{Aoki:2020prb}. 
Indeed the scalar curvature and Kretschmann scalar behave respectively as
\begin{eqnarray}
	\label{scalars2}
	R&=&\left\{
\begin{array}{ll}
  2(A-B-C), & x(t) \le 1 ,    \\
  \\
  0, &  x(t) > 1 ,   \\
\end{array}
\right.
\\
	\label{krets2}
R_{\mu\nu\sigma\rho}R^{\mu\nu\sigma\rho} &=&
\left\{
\begin{array}{ll}
  \dfrac{16e^{2\omega t}}{r^4}+\cdots, & x(t) \le 1,     \\
  \\
  \dfrac{12 h^2}{r^6}, &  x(t) > 1,    \\
\end{array}
\right.
\end{eqnarray}
which shows a singularity at $r=0$ in Kretschmann scalar is weaker than that of the Schwarzschild BH for a finite $t$, thank to the property of the mass function that $m(0,t)=0$. Since the region of $r$ satisfying $x(t)\le 1$ shrink to $r=0$ as $t\to\infty$, 
$R=0$  ($r \not= 0$) and $R_{\mu\nu\sigma\rho}R^{\mu\nu\sigma\rho} =12 h^2 r^{-6}$ ($r \not= 0$) in this limit.

The mass function Eq.~\eqref{eq:SBH_mass} generates non-zero energy momentum tensor  only at $r=0$ in terms of the delta function~\cite{Aoki:2020prb}, which agrees with the result by the distributional method~\cite{Balasin:1993fn,Balasin:1993kf}.
The total energy is evaluated simply as
\begin{eqnarray}
E &=& E_{\rm matter} ={8\pi\over \kappa}\int_0^\infty dr \partial_r m(r,t)\nonumber \\
&=&  {8\pi\over \kappa}\left[ m(\infty,t) -m(0,t)\right] =\frac{\cal M}{G_N},
\label{eq:BH2_energy}
\end{eqnarray}
which is manifestly conserved, where we use $m(\infty,t)={\cal M}$ and  $m(0,t)=0$ for all $t$.
Eq.~\eqref{eq:BH2_energy} in the $t\to\infty$ limit is interpreted that the total energy (mass) of the Schwarzschild BH
can be evaluated from the energy momentum tensor obtained through the mass function Eq.~\eqref{eq:SBH_mass}~\cite{Aoki:2020prb}.

As far as the energy conditions are concerned, while the WEC holds for $\alpha \le 1$,
NEC and SEC are satisfied for $\alpha \le \alpha_0\simeq 3.13422$. 
As in the case before, the DEC is always violated even in the static case ($\alpha=0$).
See the appendix~\ref{sec:EnergyCondition} for derivations.

We  conclude that the model II provides a physically reasonable analytic model for gravitational collapse toward the  Schwarzschild black hole,
which may reveal a nature of its singularity at $r=0$.

\section{Conclusion}
\label{con}

\noindent 
We have seen that integrable BHs are an excellent alternative to explore the inner geometry beyond the event horizon, not only in the static case, but also in the dynamical one.
In this regards, describing gravitational collapse inside BHs without resorting to numerical gravity could seem a chimera. However, in this work we show that an exact analytic analysis is indeed possible. In particular, we provide a description of the last stage of spherically symmetric collapse giving rise to the Schwarzschild solution.

Understanding the formation or existence of spacetime singularities is vital not only to clearly determine the limitations of GR, but also the viability of its possible extensions. In this matter, the energy conditions are a reasonable guide for analyzing this subject~\cite{Hawking:1973uf}. Their violation may indicate that GR has already surpassed its maximum domain, something that could happen in environments of extreme curvature.
The results of our work show that these conditions continue to be a good guide, or rather, GR is perfectly capable of describing gravitational collapse within the most extreme environment, and doing it without resorting to exotic matter.
More interestingly, the gravitational collapse in the second model leads to Eq.~\eqref{eq:SBH_mass} as
the  metric of the Schwarzschild BH in a distributional sense~\cite{Balasin:1993fn, Aoki:2020prb}.

\subsection*{Acknowledgments}
\vspace*{1mm}
JO thanks YITP for all their support during the development of this research work. This work is partially supported by ANID
FONDECYT Grant No. 1210041.
SA is supported in part by  the Grant-in-Aid of the Japanese Ministry of Education, Sports, Culture, Sciences and Technology (MEXT) for Scientific Research (No.~JP22H00129). 
\appendix
\section{Energy conditions}
\label{sec:EnergyCondition}
Since the energy momentum tensor is not diagonal, the standard criteria for  various energy conditions are not directly applicable.{\footnote{{
 Conditions obtained in this appendix agree with those from a general method in Refs.~\cite{Maeda:2018hqu,Maeda:2022vld}. }  }
In this appendix, we explicitly derive various energy conditions for such an energy momentum tensor, which is written as
\begin{eqnarray}
T^{\mu\nu} &=& \left(A-{B\over 2}\right) e_0^\mu e^\nu_0 +\left({3B\over 2} - A\right) e^\mu_1 e^\nu_1 \nonumber \\
& -& {B\over 2} ( e^\mu_0 e^\nu_1 + e_1^\mu e^\nu_0) 
- C\sum_{i=1}^2 e^\mu_i e^\nu_i,
\end{eqnarray}
 where $e^\mu_i$ satisfies $g_{\mu\nu} e^\mu_i e^\nu_j =\eta_{ij}={\rm diag}(-1,1,1,1)$ for $i,j=0,1,2, 3$.
Using a future-directed vector $v^\mu :=\sum_{i=0}^3 a_i e^\mu_i$, whose norm is given by $v^\mu v_\mu = -c_0^2$ with $c_0=1$ for a normalized time-like vector or $c_0=0$ for a null vector, we have
\begin{eqnarray}
T_{\mu\nu} v^\mu v^\nu =\left( A-{B\over 2} -C\right) R^2 + B R^2 f(s) + C c_0^2,~~~~
\label{eq:Tvv}
\end{eqnarray}
where $f(s)=s^2+s \sqrt{s^2+1}$, and $a_i$'s are parametrized as
\begin{eqnarray}
a_0= R\sqrt{1+s^2}, \ a_1= R s, \ R=\sqrt{c_0^2+a_2^2+a_3^2} ~~~~ 
\end{eqnarray}
with $c_0\le R<\infty$ and $-\infty < s <\infty$. Since $f(s)$ monotonically increases with $\lim_{s\to -\infty} f(s)=-1/2$ and
$B= 2 \dot{m}/\kappa r^2 \ge 0$ for a gravitational collapse, we have a lower bound as
\begin{eqnarray}
T_{\mu\nu} v^\mu v^\nu \ge \left( A-B-C\right) R^2  + C c_0^2.~~~~
\end{eqnarray}

Let us consider various energy conditions in terms of $A,B,C$.
\begin{description}
\item[ NEC]  The null energy condition is given for $c_0=0$ as
\begin{eqnarray}
\left( A-B-C\right) R^2  \ge 0 \Rightarrow A-B-C \ge 0.
\label{eq:NEC}
\end{eqnarray}

 \item[WEC] The weak energy condition is obtained for $c_0=1$ as
 \begin{eqnarray}
 \left( A-B-C\right) R^2  + C \ge 0
 \end{eqnarray}
with $c_0 \le R < \infty$. Therefore the necessary and sufficient condition becomes
\begin{eqnarray}
A-B-C \ge 0, \qquad A-B \ge 0.
\label{eq:WEC}
\end{eqnarray}

\item[SEC] The strong energy condition reads
\begin{eqnarray}
\left(T_{\mu\nu} -{1\over 2} T g_{\mu\nu} \right)v^\mu v^\nu \ge 0, \ T=2(B-A-C),~~~
\end{eqnarray}
which implies
\begin{eqnarray}
(A-B-C) R^2 + B-A \ge 0
\end{eqnarray}
for $ 1\le R< \infty$. Therefore we need
\begin{eqnarray}
A-B-C \ge 0, \qquad C \le 0.
\label{eq:SEC}
\end{eqnarray}

\item[DEC] The dominant energy condition requires that                                                                                                                                                                                                                                                                                                                                                                         $-T^\mu{}_\nu v^\nu$ 
must be a future directed time-like or null field for an arbitrary future directed unit normal time-like vector $v^\mu$.  This condition  reads
\begin{eqnarray}
g_{\mu\nu} T^\mu{}_\alpha v^\alpha T^\nu{}_\beta v^\beta &=& R^2 (B-A) \left[ A+2B f(s)\right]\nonumber \\
& +&  C^2(R^2-1) \le 0
\end{eqnarray}
for $1\le R<\infty$ and $-\infty < s < \infty$, where $f(s)$ is defined before.
In the $s\to-\infty$ limit, we obtain a necessary condition 
\begin{eqnarray}
\left[C^2-(B-A)^2\right] {R^2} - C^2 \le 0, 
\end{eqnarray}
which leads to
\begin{eqnarray}
C^2-(B-A)^2\le 0,  
\label{eq:DEC1}
\end{eqnarray}
while the $s\to \infty$ limit implies
\begin{eqnarray}
2B (B-A) \le 0 \Rightarrow B-A \le 0,
\end{eqnarray}
where we use a fact that $B\ge 0$ for gravitational collapse to obtain the last inequality. 
\end{description}

\subsection{Energy conditions for the model I}
 The model I with $m(r) =r-r^3/h^2 + r^4/2h^3$ in Eq.~\eqref{M2} leads to
\begin{eqnarray}
&&A-B = {e^{-\omega t}(1-x)^2\over \kappa r^2}\left[-b_0(x) \alpha + c_0(x)\right], \\
&&C  = -{e^{-\omega t}x(1-x)\over \kappa r^2} \left[ a_1(x)\alpha^2 - b_1(x) \alpha + c_1(x) \right], \\
&&A-B-C = {e^{-\omega t}(1-x)\over 2\kappa r^2}\left[ a_2(x)\alpha^2 - b_2(x) \alpha + c_2(x) \right], \nonumber \\
\end{eqnarray}
where $0\le x:= r/h\le 1 $, $0\le \alpha = h\omega$, and
\begin{eqnarray}
b_0(x)&=& (1-x)(1+x), \
c_0(x) = 2(1+2x), \nonumber\\
a_1(x)&=& (1-x)^2(1+x), \
b_1(x) = 4(1-x)(1+2x),\nonumber \\
c_1(x) &=&12 x, \ 
 a_2(x) = x(1-x)^2(1+x),
\nonumber \\
b_2(x)&=&2(1-x)(3x^2+2x+1), \
c_2(x)=4(x^2+x+1). \nonumber \\
\end{eqnarray}

\begin{description}
\item[NEC]  The NEC requires $A-B-C \ge 0$, which leads to
\begin{eqnarray}
\alpha \le \alpha^{\rm NEC}:= \min_{0\le x\le 1} \alpha_-(x),
\end{eqnarray}
where $\alpha_-$ is a smaller solution to $A-B-C=0$, given by $\alpha_-=(b_2-\sqrt{b_2^2-4a_2c_2})/2a_2$.
Explicitly, we obtain
\begin{eqnarray}
\alpha^{\rm NEC} =\alpha_-(0) = 2.
\end{eqnarray}

\item[WEC] An additional condition $A-B \ge 0$ for the WEC is easily solved as
\begin{eqnarray}
\alpha \le \alpha^{\rm WEC}= \min_{0\le x\le 1} {c_0(x)\over b_0(x)} =  {c_0(0)\over b_0(0)} =2. 
\end{eqnarray}
Therefore the WEC requires
\begin{eqnarray}
\alpha \le \min(\alpha^{\rm NEC}, \alpha^{\rm WEC})  = 2.
\end{eqnarray}

\item[SEC] An additional condition $C\le 0$ for the SEC is solved similarly as
 \begin{eqnarray}
\alpha \le \alpha^{\rm SEC}:= \min_{0\le x\le 1} \alpha_-(x)=\alpha_-(0)=0,
\end{eqnarray}
where $\alpha_-=(b_1-\sqrt{b_1^2-4a_1c_1})/2a_1$. Therefore the SEC is never satisfied except the static case ($\alpha=0$).

\item[DEC] Since $B-A \sim (1-x)^2$ but $C\sim (1-x)$, the condition eq.~\eqref{eq:DEC1} can not be satisfied near $x=1$.
Therefore, the DEC is always violated in all cases including the static one.
\end{description}

\subsection{Energy conditions for the model II}
In the model II with Eq.~\eqref{M2}, we have
\begin{eqnarray}
&&A-B = {2(1-x)^2(2x+1)\over \kappa r^2}\left[-b_0(x) \alpha + c_0(x,t)\right], \nonumber \\
&&C  = -{x(1-x)\over \kappa r^2} \left[ a_1(x,t)\alpha^2 - b_1(x) \alpha + c_1(x,t) \right], \nonumber\\
&&A-B-C = {(1-x)\over 2\kappa r^2}\left[ a_2(x,t)\alpha^2 - b_2(x) \alpha + c_2(x,t) \right], 
\nonumber \\
\end{eqnarray}
where
\begin{eqnarray}
b_0&=&x, \
c_0 = e^{\omega t}, 
a_1=e^{-\omega t} x {b_1\over 2} , \nonumber \\
b_1 &=& 2(8x^2-x-1),\ c_1 =6 e^{\omega t} x,  
\nonumber \\
 a_2 &=& x a_1,\
b_2=12x^3, \
c_2=2e^{\omega t}(x^2+x+1).\nonumber \\
\end{eqnarray}

\begin{description}
\item[NEC]  The NEC requires $A-B-C \ge 0$, which leads to
\begin{eqnarray}
\alpha \le \alpha^{\rm NEC}:= \min_{0\le x\le 1} \alpha_-(x,0)=3.13422,
\end{eqnarray}
where $\alpha_-=(b_2-\sqrt{b_2^2-4a_2c_2})/2a_2$,
and the minimum value is calculated by Mathematica.

\item[WEC] An additional condition $A-B \ge 0$ for the WEC is easily solved as
\begin{eqnarray}
\alpha \le \alpha^{\rm WEC}= \min_{0\le x\le 1} {c_0(x,0)\over b_0(x)} =  {c_0(1,0)\over b_0(1)} =1. 
\end{eqnarray}
Therefore the WEC requires
\begin{eqnarray}
\alpha \le \min(\alpha^{\rm NEC}, \alpha^{\rm WEC})  = 1.
\end{eqnarray}

\item[SEC] An additional condition $C\le 0$ for the SEC is solved similarly as
 \begin{eqnarray}
\alpha \le \alpha^{\rm SEC}:= \min_{0\le x\le 1} \alpha_-(x,0)=7.74262,
\end{eqnarray}
where $\alpha_-=(b_1-\sqrt{b_1^2-4a_1c_1})/2a_1$, and  the minimum value is calculated by Mathematica.
Therefore the SEC 
requires
\begin{eqnarray}
\alpha \le \min(\alpha^{\rm NEC}, \alpha^{\rm SEC})  = 3.13422. 
\end{eqnarray}

\item[DEC] Since $B-A \sim (1-x)^2$ but $C\sim (1-x)$, the condition eq.~\eqref{eq:DEC1} can not be satisfied near $x=1$.
Therefore, the DEC is always violated in all cases including the static one.
\end{description}

\subsection{Summary}
We summarize the requirement by various energy conditions for $\alpha =h\omega$, which controls a speed of gravitational collapse  
in Table~\ref{tab:EC}.

\begin{table}[bht]
	\caption{\label{tab:EC} The allowed range of $\alpha= h\omega \ge 0$ for  various energy conditions.
	The symbol $-$ means no allowed range exists, and $\alpha_0=3.13422$. }
		\begin{tabular}{| c | c| c| c| c| }
		\hline
			Type of model & NEC & WEC & SEC & DEC 
			\\  \hline
			Type I & $\alpha \le 2$ & $\alpha \le 2$ & $\alpha =0$ & $-$  \\ \hline
			Type II & $\alpha \le \alpha_0$ & $\alpha \le 1$ & $\alpha \le \alpha_0$ & $-$  \\ \hline
		\end{tabular}
\end{table}
 
%
%
%
\bibliography{references.bib}
\bibliographystyle{apsrev4-1.bst}
%
%
\end{document}